\documentclass[submitting]{nst}
\usepackage{subfigure,dcolumn}
\usepackage[T2A,T1]{fontenc}
\usepackage[russian,english]{babel}
\usepackage{listings}
\usepackage{multirow}
\begin{document}
\title{Development and Performance Study of a Capillary Liquid Scintillator Neutron Detector}
\author{Guang Luo}
\affiliation{Center for Intense Laser Application Technology, and College of Engineering Physics, Shenzhen Technology University, Shenzhen 518118, China}
\author{Jian Yu}
\affiliation{Center for Intense Laser Application Technology, and College of Engineering Physics, Shenzhen Technology University, Shenzhen 518118, China}
\author{Dikai Li}
\affiliation{Center for Intense Laser Application Technology, and College of Engineering Physics, Shenzhen Technology University, Shenzhen 518118, China}
\author{Yanmeng Dai}
\affiliation{Center for Intense Laser Application Technology, and College of Engineering Physics, Shenzhen Technology University, Shenzhen 518118, China}
\author{Meiling Chen}
\affiliation{Center for Intense Laser Application Technology, and College of Engineering Physics, Shenzhen Technology University, Shenzhen 518118, China}
\author{Jiaqiang Zou}
\affiliation{Center for Intense Laser Application Technology, and College of Engineering Physics, Shenzhen Technology University, Shenzhen 518118, China}
\author{Ke Yan}
\affiliation{Center for Intense Laser Application Technology, and College of Engineering Physics, Shenzhen Technology University, Shenzhen 518118, China}
\author{Shaoxuan Cui}
\affiliation{Center for Intense Laser Application Technology, and College of Engineering Physics, Shenzhen Technology University, Shenzhen 518118, China}
\author{Ge Jin}
\affiliation{North Night Vision Technology (Nanjing) Research Institute Co., Ltd., Nanjing 210000, China}
\author{Zhao Xu}
\affiliation{North Night Vision Technology (Nanjing) Research Institute Co., Ltd., Nanjing 210000, China}
\author{Chunhui Zhang\footnotemark[2]}
\affiliation{School of Nuclear Science and Technology, Lanzhou University, Lanzhou 730000, China}
\author{Leifeng Cao\footnotemark[1]}
\affiliation{Center for Intense Laser Application Technology, and College of Engineering Physics, Shenzhen Technology University, Shenzhen 518118, China}

\footnotetext[1]{Corresponding author: \href{mailto:caoleifeng@sztu.edu.cn}{caoleifeng@sztu.edu.cn}}
\footnotetext[2]{Corresponding author: \href{mailto:zhangchunhui@lzu.edu.cn}{zhangchunhui@lzu.edu.cn}}

\begin{abstract}
Capillary liquid scintillator detectors are promising for high-resolution neutron imaging, yet experimental data on their light spread mechanism and spatial performance remain limited. Here, we report a neutron detector based on a hexagonal capillary array filled with EJ-309 liquid scintillator, with an inner diameter of $\sim$50~$\mu$m and a camera readout of 9~$\mu$m pixels. Laser experiments show that the FWHM of the full light spot decreases from 260~$\mu$m to 90~$\mu$m with a metal light absorber, confirming effective suppression of lateral light spread. Using an AmBe neutron source, an effective field of view with a $5\sigma$ threshold was established from background frames. For single-capillary events, the pulse height spectrum follows a Landau distribution with a most probable value of $0.133 \pm 0.001$ (stat.), and the intrinsic detection efficiency is $10.07\% \pm 1.26\%$ (stat.) $\pm 1.43\%$ (syst.), corresponding to about 13.55\% when normalized to the active liquid scintillator area. The point spread function core yields a radial FWHM of 12.8~$\mu$m and a centroid positioning precision of approximately 5.5~$\mu$m (1$\sigma$), while the intrinsic position resolution is limited by the capillary pitch to 54~$\mu$m. Linearity is good for $1\rightarrow2$ capillaries, with deviation appearing for $2\rightarrow3$ capillaries due to additional capture of spread light. These results provide experimental basis and physical understanding for imaging applications of capillary liquid scintillator neutron detectors.
\end{abstract}

%\textcolor{red}{This article introduces the unique design of the module and reports the excellent performance of all modules, providing guidance and important reference for the process design of scintillation detectors with WLS-fibers.}


\maketitle

\section{Introduction}
\label{sec:intro}
High spatial resolution neutron detection is a critical requirement for inertial confinement fusion (ICF) implosion diagnostics\cite{Frenje2020}. During implosion, 14 MeV neutrons carry spatial information about the burning region. Position-sensitive measurements via imaging systems allow the reconstruction of implosion asymmetry and fuel mixing\cite{Caillaud2012a,Caillaud2012b,Miyanaga1990,He2022}. In recent years, as laser fusion facilities advance toward ignition, increasingly stringent demands have been placed on the spatial resolution and dynamic range of neutron imaging detectors\cite{Guler2012,Isabelle2010,Wilke2008,Ma2025}.

Among various technical approaches, capillary liquid scintillator detectors have attracted considerable attention due to their unique structural advantages. Compared with bubble detectors\cite{Lerche2003}, they offer higher detection efficiency; compared with plastic scintillator fiber detectors\cite{Mor2012,Zhang2023,Li2022,Chen2022}, they provide better spatial resolution potential. Their basic structure consists of a liquid scintillator filling a micrometer-scale capillary array. Scintillation light produced by neutron interactions is read out by an optical imaging system\cite{Disdier2004,Disdier2006}. Since each capillary can be regarded as an independent pixel unit, the intrinsic spatial resolution of the detector is primarily determined by the capillary pitch. Previous studies on capillary detectors for ICF neutron imaging have demonstrated that spatial resolution can be progressively improved by reducing the capillary diameter, enhancing light collection efficiency, and replacing hydrogen with deuterium in the liquid scintillator, achieving a spatial resolution of 325 $\mu$m for 14 MeV neutrons\cite{Disdier2004,Zhao2023,Ding2024}. In terms of aperture design, coded apertures and penumbral apertures have also been extensively optimized\cite{Chen2019,Vogel2014}. A previous Monte Carlo simulation study on an ideal capillary liquid scintillator detector for 14 MeV fusion neutrons~\cite{Zhang2022} predicted its performance under ideal conditions. Other studies have introduced pattern recognition algorithms~\cite{Song2018} and optimized Hough transforms to improve the accuracy of neutron position determination~\cite{Li1989a,Fernandes2008,He2019,Zhang2022}. In addition, fast neutron imaging spectrometers based on liquid scintillator capillaries and Monte Carlo simulations of glass scintillator microfibers have been reported\cite{Song2016,Song2020,Kang2025,Liu2023}.

However, converting the above simulation predictions into practical detection capability still requires overcoming a key physical bottleneck: lateral optical spread of scintillation light between capillary walls. Ideal simulations typically assume that scintillation light is perfectly confined within a single capillary. In a real detector, however, photons can spread into adjacent capillaries through wall refraction, reflection, and crosstalk. This spread not only degrades the energy resolution of single-capillary events but also affects the charge reconstruction accuracy of multi-capillary overlapping events. Although previous studies have attempted to improve performance by reducing the aperture and using deuterated scintillators\cite{Disdier2004}, direct experimental quantification of the light spread mechanism and active suppression of spread through physical means remain relatively limited. As a widely used liquid scintillator, EJ-309 has been systematically studied for its neutron light output response and resolution functions\cite{Enqvist2013,Norsworthy2018}, and its ionization quenching effect and light output response have also been investigated through systematic experiments and semi-empirical calculations\cite{Tretyak2010,Swiderski2012}. However, experimental studies on its light spread behavior in capillary arrays are still limited.

To address these issues, this paper presents a systematic experimental study on the light spread mechanism and performance of a capillary liquid scintillator neutron detector. The remainder of this paper is organized as follows. Section~\ref{sec:detector} describes the detector structure and fabrication process. Section~\ref{sec:laser} presents the laser experiments on light spread with and without a metal light absorber. Section~\ref{sec:neutron} reports the neutron irradiation experiments using an AmBe source, including background treatment, event identification, detection efficiency, single-capillary pulse height spectrum, linearity analysis, and positioning performance. Finally, Section~\ref{sec:summary} summarizes the main results and gives an outlook.

\section{Detector Structure and Fabrication}
\label{sec:detector}

The capillary array was developed in collaboration with North Night Vision Technology (Nanjing) Research Institute Co., Ltd.\cite{NorthNightVision}. The fabrication process is briefly described as follows. First, glass powder is prepared and melted, then cast into a glass tube. The tube is then drawn into single fibers, which are arranged into multifiber bundles. These bundles are further drawn into multifibers, which are finally arranged into a screen and pressed to form the capillary array.
\begin{figure}[htbp]
   \subfigure[]{
   \begin{minipage}[t]{0.54\linewidth}
   \centering
   \includegraphics[width=4.9cm]{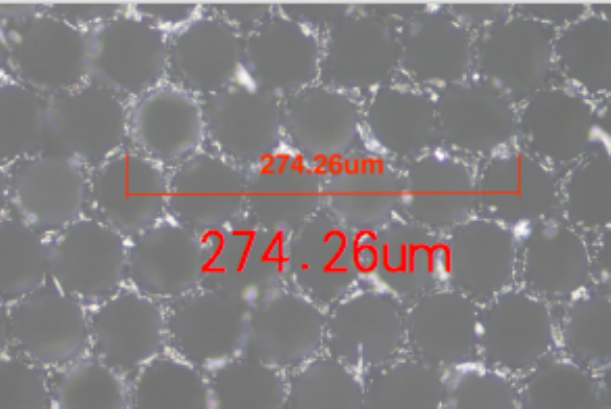}
    \label{fig:Capillary_Microscopy}
    \end{minipage}
    }
    \subfigure[]{
    \begin{minipage}[t]{0.35\linewidth}
    \centering
    \includegraphics[width=2.8cm]{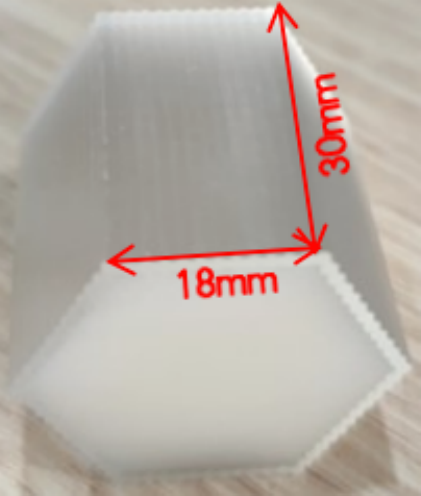}
    \label{fig:Finished_product_photo}
    \end{minipage}
    }
    \caption{(a) The microscopic image shows the hexagonal capillary array. (b) The photograph shows the finished quartz capillary array.}
    \label{1}
\end{figure}
The resulting capillary array has a hexagonal close-packed structure. Each capillary has a regular hexagonal cross-section with an inner diameter (across flats) of 50~$\mu$m. The material is quartz glass with a refractive index of 1.5. The wall thickness is approximately 3--5~$\mu$m. Figure~\ref{1} shows (a) a microscopic image of the capillary array and (b) a photograph of the finished quartz capillary array. The finished array is a hexagonal prism with a distance across flats of 18~mm and a length of 30~mm.

The capillary array was then filled with EJ-309 liquid scintillator. The refractive index of EJ-309 is 1.57, which is higher than that of the quartz glass (1.5). This refractive index difference causes total internal reflection at the interface, helping to confine the scintillation light within each capillary. The filling and sealing procedures are described as follows.

\textbf{Liquid scintillator filling:} Taking advantage of the capillary effect, the capillary array was immersed in EJ-309 liquid scintillator in a nitrogen atmosphere. The scintillator was drawn into the capillaries by capillary action. After filling, the array was kept in a nitrogen atmosphere at low temperature.

\textbf{Capillary array sealing:} The two end faces of the capillary array were rapidly sealed with optical adhesive EJ-500 in a nitrogen atmosphere at low temperature. A reflective film was placed on one end face to enhance light collection.

\begin{figure}[htbp]
    \centering
    \includegraphics[width=0.95\linewidth]{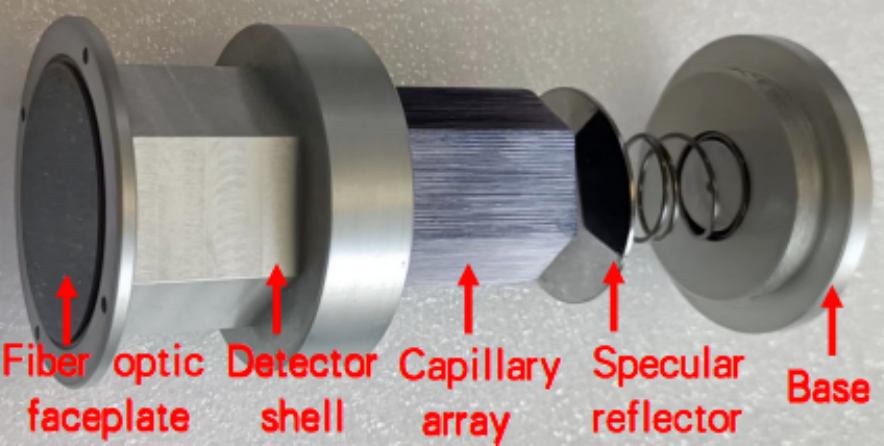}
    \caption{Photograph of the fully assembled detector, showing the fiber optic faceplate, detector shell, capillary array, specular reflector, and base.}
    \label{fig:assembly}
\end{figure}
After filling and sealing, the capillary array was assembled into the detector housing. Figure~\ref{fig:assembly} shows the fully assembled detector. The housing consists of a base, a specular reflector, a detector shell, a capillary array, and a fiber optic faceplate. The capillary array is mounted inside the detector shell. The specular reflector is placed at the rear end of the capillary array to reflect backward-traveling scintillation light towards the readout end. The fiber optic faceplate is located at the front end to couple the scintillation light to the readout camera. The spring washer provided uniform pressure to hold the capillary array in position. The base closes the rear end of the housing. All components were assembled in a nitrogen atmosphere to avoid moisture and oxygen contamination.

To investigate the lateral light spread between capillaries, two types of detector samples were prepared: one with a metal light absorber coated on the outer walls of the capillaries, and one without. The effect of the metal light absorber on light spread was then studied by laser experiments, which are presented in the next section.
\section{laser experiments}
\label{sec:laser}
\begin{figure}[htbp]
   \subfigure[]{
   \begin{minipage}[t]{0.46\linewidth}
   \centering
   \includegraphics[width=4.1cm]{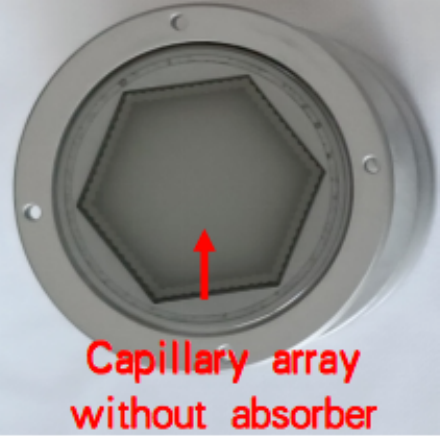}
    \label{fig:absorber_compare_a}
    \end{minipage}
    }
    \subfigure[]{
    \begin{minipage}[t]{0.44\linewidth}
    \centering
    \includegraphics[width=3.8cm]{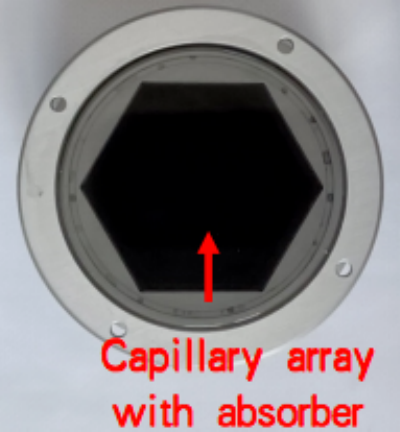}
    \label{fig:absorber_compare_b}
    \end{minipage}
    }
    \caption{(a) The photograph shows the capillary array without a light absorber. (b) The photograph shows the capillary array with a light absorber. The array with the absorber appears visibly darker, indicating effective suppression of light leakage.}
    \label{fig:absorber_compare}
\end{figure}
Two identical capillary arrays were prepared: one without any light absorber, and one with a light absorber coated on the outer walls of the capillaries. The purpose of the light absorber is to suppress the lateral transmission of scintillation light between adjacent capillaries. In this work, iron oxide was used as the absorber material.

Figure~\ref{fig:absorber_compare} shows photographs of the two detectors. The detector without the absorber is shown in Fig.~\ref{fig:absorber_compare}(a), where the capillary array appears relatively bright and semi-transparent. In contrast, the detector with the absorber is shown in Fig.~\ref{fig:absorber_compare}(b), where the capillary array is visibly darker. This visual contrast directly indicates that the light absorber effectively absorbs light that escapes from the capillary walls, thereby reducing light leakage and suppressing lateral light spread.
\begin{figure}[htbp]
    \centering
    \includegraphics[width=0.95\linewidth]{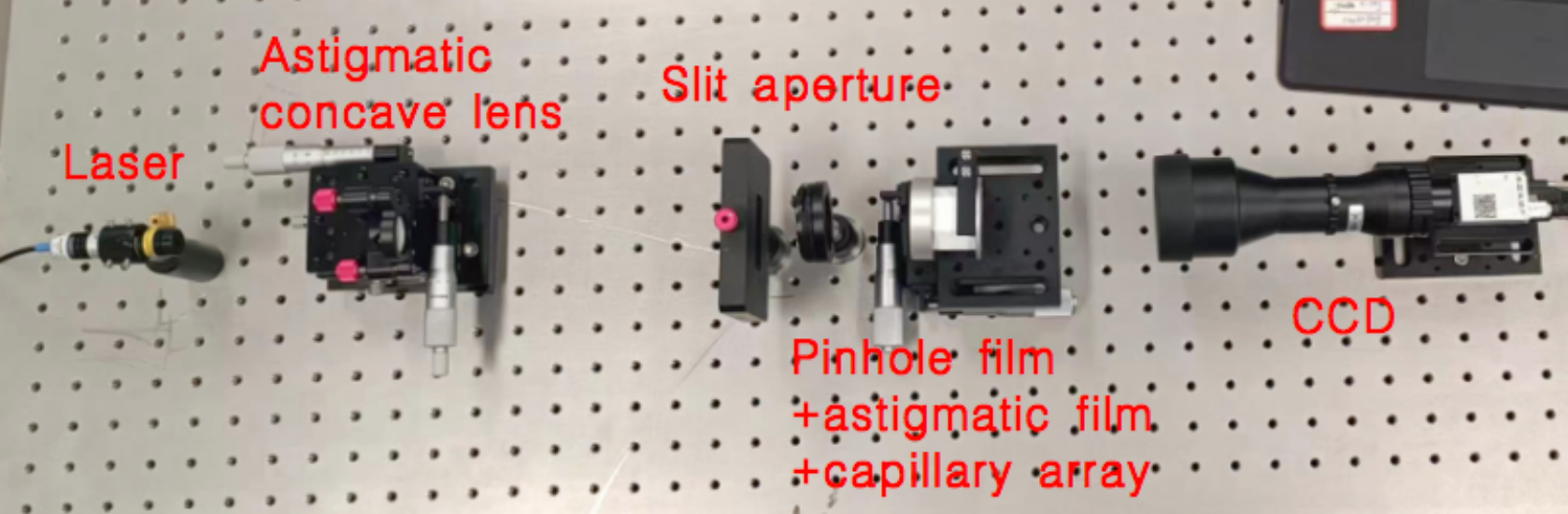}
    \caption{Photograph of the laser experimental setup, including the laser, astigmatic concave lens, slit aperture, pinhole film, astigmatic film, capillary array, and CCD camera.}
    \label{fig:laser_setup}
\end{figure}
To quantitatively characterize the suppression effect of the absorber, A laser experiment was designed to simulate a point-like light source on the capillary array, in order to study the lateral light spread between capillaries. Figure~\ref{fig:laser_setup} shows a photograph of the experimental setup. The setup consists of a laser, an astigmatic concave lens, a slit aperture, a pinhole film, an astigmatic film, the capillary array, and a CCD camera.

A 532 nm green laser was chosen as the light source. This wavelength selection is based on the following considerations. The emission peak of EJ-309 liquid scintillator is around 425 nm. Since 532 nm is far above the absorption band of EJ-309\cite{Enqvist2013,Norsworthy2018}, the green laser does not excite fluorescence re-emission in the liquid scintillator. Therefore, the laser light undergoes purely optical transmission (refraction, reflection, and scattering) within the capillary array. This allows the lateral light spread to be studied without the interference of the scintillation emission process. Although the wavelength differs from the actual scintillation light (425 nm), the absorption and reflection characteristics of the metal light absorber in the visible range vary smoothly with wavelength. Hence, the 532 nm laser experiment can provide a valid qualitative comparison of the light spread behavior with and without the absorber.

In the setup, the laser beam was first expanded by the astigmatic concave lens. The beam then passed through the slit aperture and the pinhole film to form an approximate point source. After passing through the astigmatic film, the light was uniformly diffused and illuminated the capillary array. The light spot at the output end was recorded by the CCD camera.
\begin{figure}[htbp]
   \subfigure[]{
   \begin{minipage}[t]{0.45\linewidth}
   \centering
   \includegraphics[width=3.8cm]{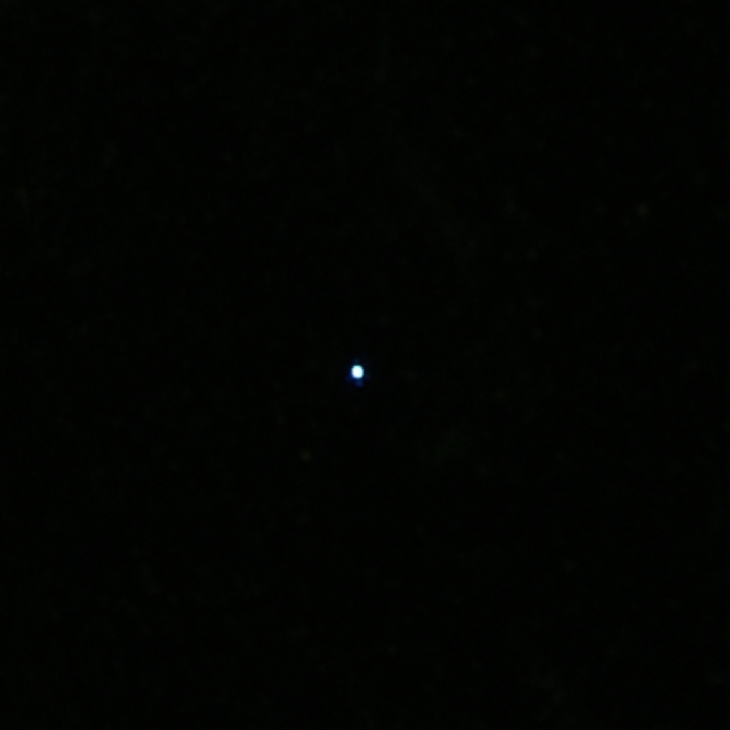}
    \label{fig:laser_with_absorber}
    \end{minipage}
    }
    \subfigure[]{
    \begin{minipage}[t]{0.45\linewidth}
    \centering
    \includegraphics[width=3.8cm]{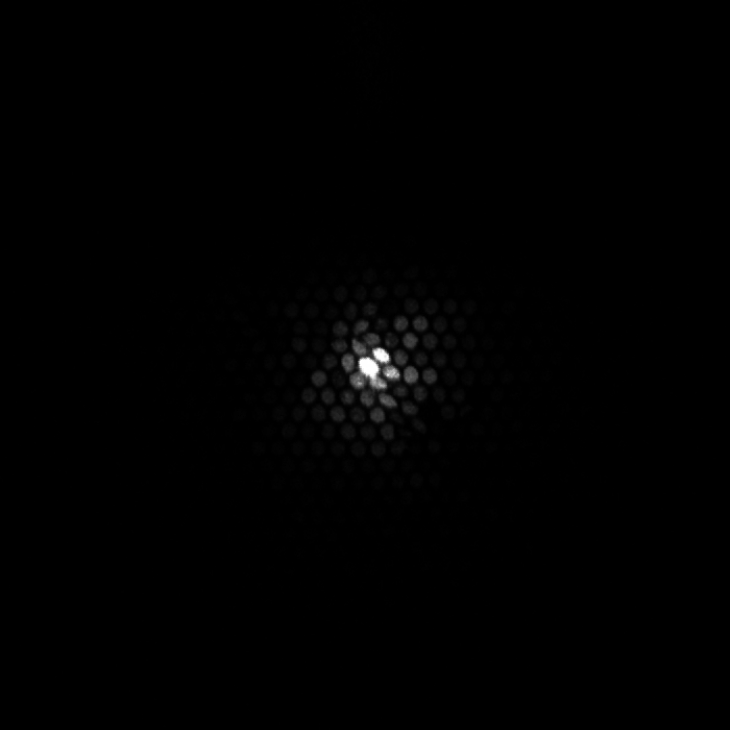}
    \label{fig:laser_without_absorber}
    \end{minipage}
    }
    \caption{Raw camera images recorded after the laser beam passed through the optical setup and illuminated the capillary array: (a) with the light absorber and (b) without the light absorber. The array with the absorber shows a single confined spot, while the array without the absorber shows a bright central spot accompanied by several secondary spots from lateral light spread.}
    \label{fig:laser_images}
\end{figure}
Figure~\ref{fig:laser_images} shows the raw camera images recorded after the laser beam passed through the optical setup and illuminated the capillary array. For the array with the light absorber, as shown in Fig.~\ref{fig:laser_images}(a), only one bright spot is observed, indicating that the light is well confined. In contrast, for the array without the absorber, as shown in Fig.~\ref{fig:laser_images}(b), a very bright central spot is accompanied by several secondary bright spots around it. These secondary spots correspond to light that has leaked out of the central region and propagated into adjacent capillaries. This visual comparison directly demonstrates that the light absorber effectively suppresses the lateral light spread.
\begin{figure}[htbp]
    \centering
    \includegraphics[width=0.95\linewidth]{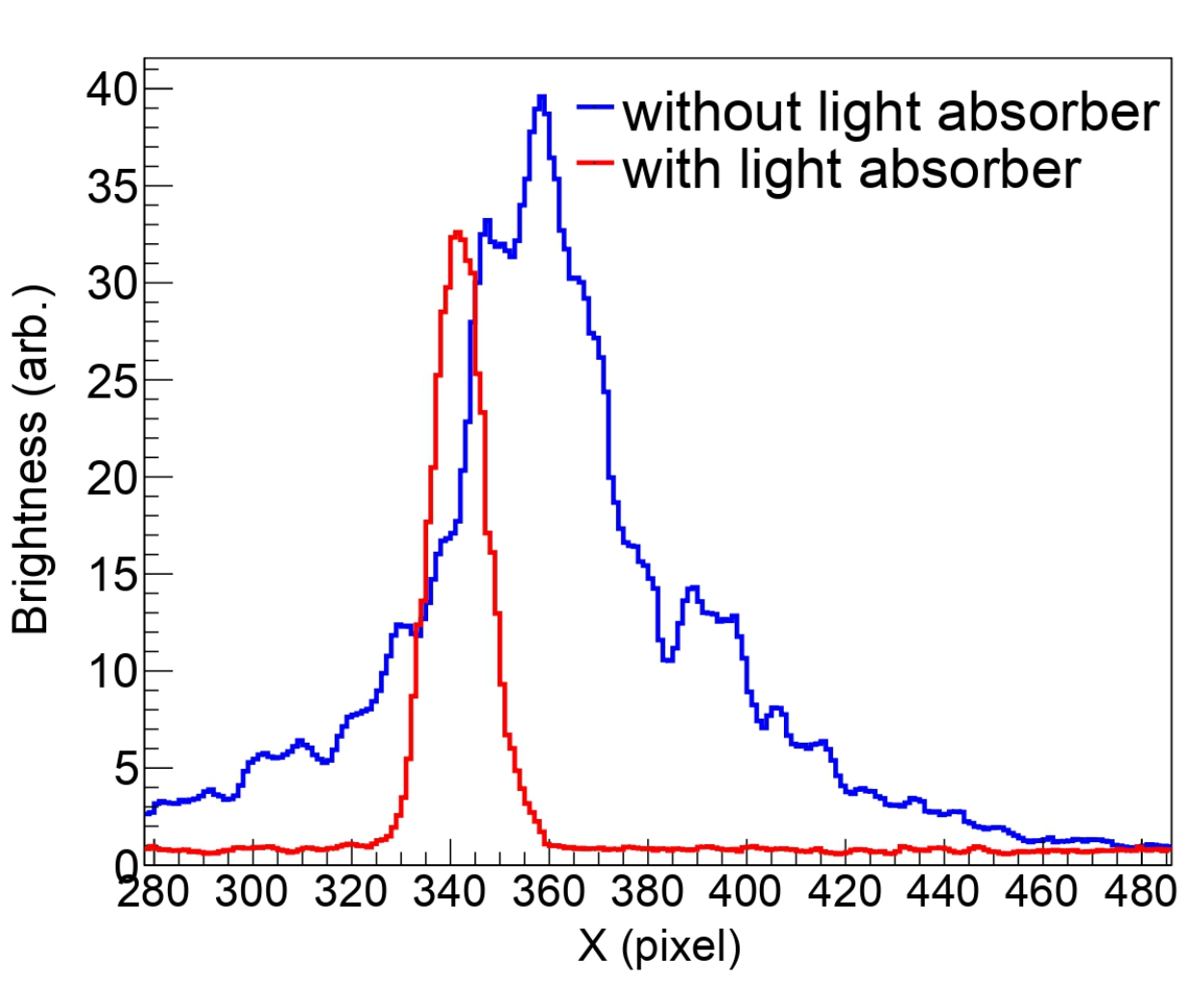}
    \caption{Brightness profiles of the light spot along the horizontal direction. The blue curve is for the array without the light absorber, and the red curve is for the array with the light absorber. The FWHM is reduced from approximately 260~$\mu$m to 90~$\mu$m when the absorber is applied.}
    \label{fig:laser_profile}
\end{figure}
To quantitatively characterize the suppression effect, the grayscale distribution of the light spot was extracted from the raw images. Figure~\ref{fig:laser_profile} shows the brightness profiles along the horizontal direction. The blue curve represents the array without the light absorber, and the red curve represents the array with the light absorber. The profile of the array without the absorber exhibits a broad main peak with a long tail and multiple secondary peaks, indicating severe lateral light spread. In contrast, the profile of the array with the absorber shows a much narrower main peak, and the secondary peaks are almost completely eliminated.

The full width at half maximum (FWHM) of the light spot was determined from these profiles. For the array without the light absorber, the FWHM is approximately 260~$\mu$m. For the array with the light absorber, the FWHM is reduced to approximately 90~$\mu$m. Given that the camera pixel size is 9~$\mu$m, these values correspond to about 29 pixels and 10 pixels, respectively. The FWHM is reduced by about 65\%, which quantitatively confirms that the metal light absorber significantly suppresses the lateral light spread between capillaries.

\section{Neutron Irradiation Experiments with an AmBe Source}
\label{sec:neutron}
In Section~\ref{sec:laser}, the laser experiments quantitatively characterized the lateral light spread in the capillary array and demonstrated that the metal light absorber reduces the light spot FWHM. However, the laser experiment only reflects the passive optical transmission properties of the capillary array. It does not directly represent the detector response to neutrons, because the scintillation light produced by neutron interactions involves additional physical processes such as ionization quenching, light generation, and isotropic emission inside the capillary. Therefore, to evaluate the overall performance of the detector in a real neutron field, a neutron irradiation experiment was carried out using an AmBe neutron source.

\subsection{Experimental Setup and Data Acquisition}
\label{subsec:neutron_setup}

\begin{figure}[htbp]
    \centering
    \includegraphics[width=0.95\linewidth]{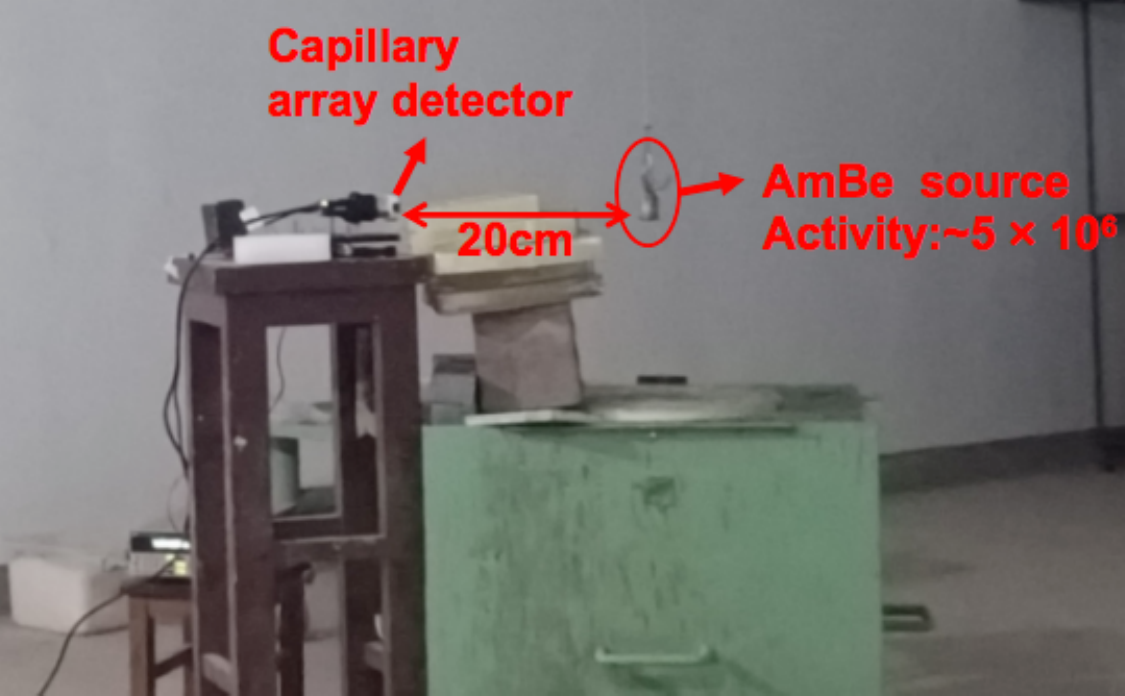}
    \caption{Photograph of the neutron irradiation experiment, showing the capillary array detector and the AmBe neutron source.}
    \label{fig:neutron_setup}
\end{figure}
The neutron irradiation experiments were performed using an AmBe neutron source with an activity of approximately $5 \times 10^6$~n/s. The source-to-detector distance was 20~cm. The exposure time was 1~s per frame. A total of 437 frames were recorded. In addition, 11 background frames were acquired with the source removed to establish the background model. Figure~\ref{fig:neutron_setup} shows a photograph of the experimental layout, including the capillary array detector and the AmBe neutron source. The camera settings, including pixel size (9~$\mu$m) and resolution (1608~$\times$~1104), were the same as those used in the laser experiments.
\subsection{Background Treatment and Effective Field of View}
\label{subsec:background}
\begin{figure*}[htbp]
    \centering
    \includegraphics[width=0.95\linewidth]{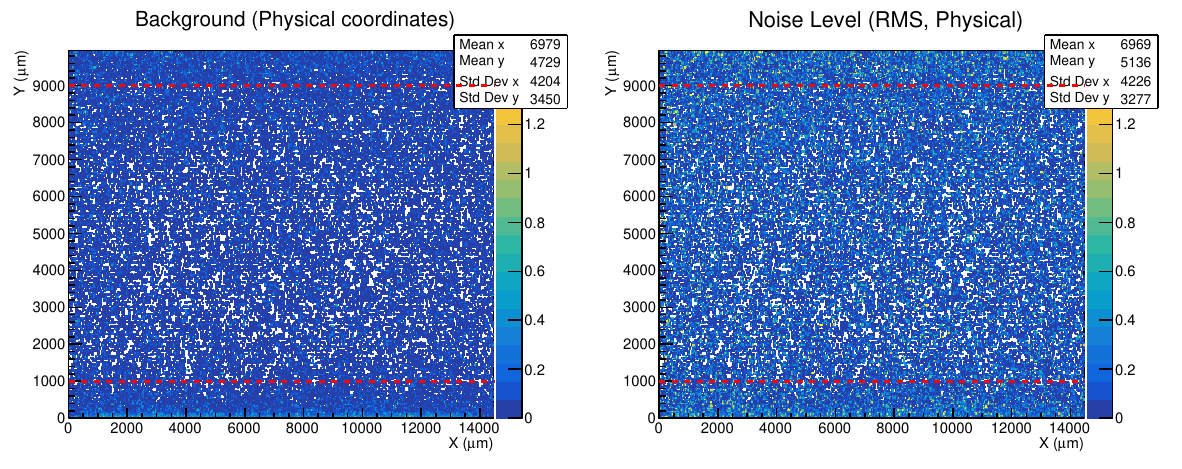}
    \caption{Background characteristics in physical coordinates: (left) mean background image and (right) RMS noise image, obtained from 11 background frames. The red dashed lines indicate the boundaries of the effective field of view ($Y = 1000~\mu$m and $Y = 9000~\mu$m).}
    \label{fig:background}
\end{figure*}
\begin{figure*}[htbp]
    \centering
    \includegraphics[width=0.95\linewidth]{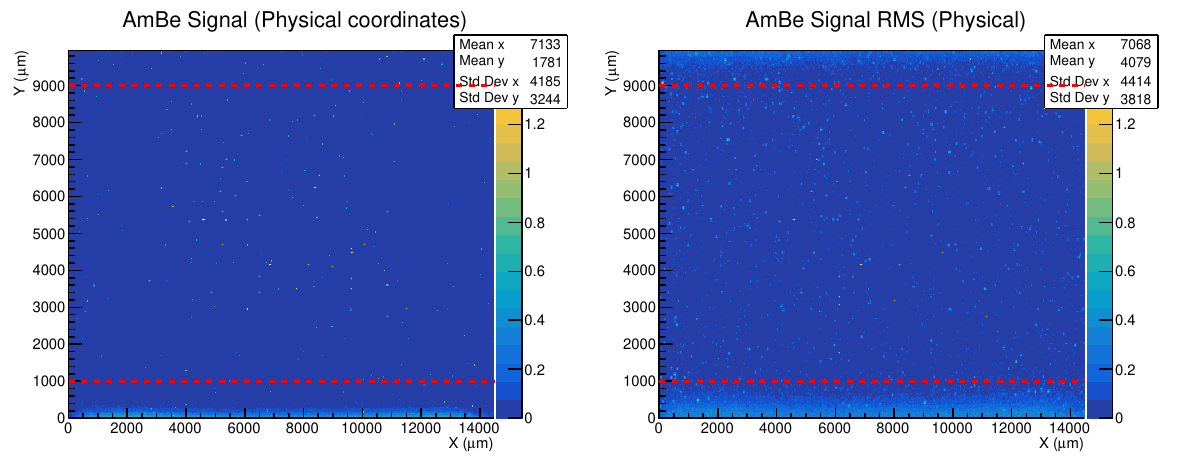}
    \caption{Average AmBe signal (left) and frame-to-frame RMS (right) from 437 frames. The red dashed lines indicate the effective field of view ($Y = 1000$--$9000~\mu$m).}
    \label{fig:ambe_mean_rms}
\end{figure*}
To establish the background model, 11 background frames were acquired with the neutron source removed. For each pixel $(i,j)$, the mean gray value $\mu_{i,j}$ and the root-mean-square (RMS) noise $\sigma_{i,j}$ were calculated over the 11 frames:
\begin{equation}
\mu_{i,j} = \frac{1}{N} \sum_{k=1}^{N} g_{k}(i,j),
\end{equation}
\begin{equation}
\sigma_{i,j} = \sqrt{\frac{1}{N} \sum_{k=1}^{N} \left[ g_{k}(i,j) - \mu_{i,j} \right]^{2}},
\end{equation}
where $g_{k}(i,j)$ is the gray value of pixel $(i,j)$ in the $k$-th background frame, and $N = 11$ is the number of background frames. The mean image $\mu$ represents the fixed-pattern background, while the RMS image $\sigma$ characterizes the pixel-wise noise level.

Figure~\ref{fig:background} shows the resulting mean background image and the RMS noise image in physical coordinates. The color scale indicates the gray value for the mean image and the noise level for the RMS image. It can be seen that both the mean background and the RMS noise are noticeably higher near the top and bottom edges of the detector (i.e., $Y < 1000~\mu$m and $Y > 9000~\mu$m). This non-uniformity may be attributed to optical vignetting and possible light leakage at the detector edges. To avoid bias from these edge effects, an effective field of view (FOV) was defined as $Y = 1000~\mu$m to $9000~\mu$m. The boundaries of this effective FOV are marked by the red dashed lines in Fig.~\ref{fig:background}. 
The average gray value and the average RMS noise within this effective FOV are found to be
\begin{equation}
\mu_{\mathrm{center}} = 0.01229, \qquad \sigma_{\mathrm{center}} = 0.01649,
\end{equation}
based on approximately $2.5 \times 10^{5}$ effective pixels in the central region. The corresponding $5\sigma$ selection threshold is then given by
\begin{equation}
\mathrm{Threshold} = \mu_{\mathrm{center}} + 5\,\sigma_{\mathrm{center}} \approx 0.09473.
\end{equation}

The choice of $5\sigma$ corresponds to a false-positive probability of approximately $5.7 \times 10^{-7}$ for a Gaussian noise distribution, i.e., about one false count in 1.7 million pixels. This ensures that the vast majority of pixels exceeding the threshold originate from genuine neutron-induced scintillation rather than background fluctuations. All subsequent event identification, efficiency calculation, and spectrum extraction are therefore performed using this $5\sigma$ threshold within the effective FOV.
The 437 AmBe signal frames were averaged and the frame-to-frame RMS was calculated. Figure~\ref{fig:ambe_mean_rms} shows the results. Both the mean and RMS are higher near the edges, confirming the need to exclude these edge regions for subsequent analysis.
\subsection{Event Identification and Classification}
\label{subsec:event_id}
After background subtraction and threshold selection, neutron-induced events were identified as connected regions of pixels exceeding the $5\sigma$ threshold. An 8-neighbor breadth-first search (BFS) algorithm was employed for connected-component labeling~\cite{Samet1986,Zhou2011}. 

Let $S = \{(i,j) \mid g(i,j) > \mathrm{Threshold}\}$ be the set of pixels above the threshold, where $g(i,j)$ is the background-subtracted gray value at pixel $(i,j)$. The 8-neighborhood of a pixel $(i,j)$ is defined as
\begin{equation}
N_8(i,j) = \left\{ (i+\delta_x, j+\delta_y) \;\middle|\; \substack{\delta_x,\delta_y \in \{-1,0,1\} \\ (\delta_x,\delta_y) \neq (0,0)} \right\}.
\end{equation}
Starting from an unvisited seed pixel $p_0 \in S$, the BFS algorithm iteratively visits all pixels in $N_8$ that belong to $S$ and have not been visited, until the queue is empty. The set of visited pixels forms a connected component $C \subset S$, which is regarded as one neutron event. 
Each connected component was then classified according to its number of pixels. Based on the capillary inner diameter of 50~$\mu$m and the camera pixel size of 9~$\mu$m, a single capillary corresponds to approximately 5--6 pixels. The classification criteria are as follows:
\begin{itemize}
    \item Single capillary (1 cap): 5--6 pixels;
    \item Two capillaries (2 caps): 7--10 pixels;
    \item Three capillaries (3 caps): 11--15 pixels;
    \item Large clusters ($>$3 caps): $\geq$16 pixels.
\end{itemize}

Events with fewer than 5 pixels were not counted as neutron candidates, as they are likely to be noise remnants or gamma-ray background events that deposit only a small fraction of their energy in the capillary. Table~\ref{tab:event_counts} summarizes the number of events in each category within the effective field of view ($Y = 1000$--$9000~\mu$m) for the 20~cm data set. A total of 437 frames were analyzed.

\begin{table}[htbp]
    \centering
    \caption{Number of events in each category within the effective field of view.}
    \label{tab:event_counts}
    \begin{tabular}{lcc}
        \hline
        Category & Pixels & Number of events \\
        \hline
        Single capillary & 5--6 & 27,476 \\
        Two capillaries & 7--10 & 16,531 \\
        Three capillaries & 11--15 & 3,826 \\
        Large clusters & $\geq$16 & 2,830 \\
        \hline
        Total ($\geq$5 pixels) & -- & 50,663 \\
        \hline
    \end{tabular}
\end{table}

It can be seen that the single-capillary events dominate the data set, accounting for the majority of all neutron candidates. This is consistent with the expectation that most neutrons interact with the scintillator in a single capillary. Two-capillary events occur when the light spot spreads into an adjacent capillary or when two neutrons hit nearby capillaries simultaneously. Three-capillary and larger events become increasingly rare. These multi-capillary events are further analyzed in Section~\ref{subsec:spectrum} to study the light spread and linearity of the detector response.
\subsection{Detection Efficiency}
\label{subsec:efficiency}
The detection efficiency was calculated by comparing the measured neutron flux rate with the theoretical flux rate. The effective area of the detector is
\begin{equation}
A_{\mathrm{eff}} = (N_x \cdot p) \times (y_{\max} - y_{\min}),
\end{equation}
where $N_x = 1608$ is the number of pixels in the $X$ direction, $p = 9~\mu$m is the pixel size, and $y_{\min} = 1000~\mu$m and $y_{\max} = 9000~\mu$m define the effective field of view. This gives $A_{\mathrm{eff}} \approx 1.158$~cm$^2$.

For a total of $M = 437$ frames with exposure time $t = 1$~s per frame, the measured neutron flux rate is
\begin{equation}
\Phi_{\mathrm{meas}} = \frac{N_{\mathrm{total}}}{A_{\mathrm{eff}} \cdot M \cdot t},
\end{equation}
where $N_{\mathrm{total}}$ is the total number of neutron events ($\geq$5 pixels) in the effective field of view.

The theoretical neutron flux rate at a distance $R$ from a point-like source of strength $S$ is
\begin{equation}
\Phi_{\mathrm{theo}} = \frac{S}{4\pi R^2}.
\end{equation}
With $S = 5 \times 10^6$~n/s and $R = 20$~cm, $\Phi_{\mathrm{theo}} \approx 199.5$~n/(cm$^2\cdot$s).

The detection efficiency is then defined as
\begin{equation}
\varepsilon = \frac{\Phi_{\mathrm{meas}}}{\Phi_{\mathrm{theo}}}.
\end{equation}

Using the $5\sigma$ threshold of 0.09474, a total of 50,663 neutron events ($\geq$5 pixels) were recorded in 437 frames within the effective field of view. The average intrinsic detection efficiency is obtained as
\begin{equation}
\varepsilon = 10.07\% \pm 1.26\%~(\mathrm{RMS}) \pm 1.43\%~(\mathrm{syst.}),
\end{equation}
where the RMS value reflects the run-to-run statistical fluctuation among the 437 frames, and the systematic error is estimated from the uncertainties in source strength ($\pm$10\%), source-to-detector distance ($\pm$5\%), and effective area ($\pm$1\%). The total uncertainty is approximately $\pm 1.90\%$.
\begin{table*}[htbp]
    \centering
    \caption{Detection efficiency and event counts under different selection thresholds.}
    \label{tab:threshold_scan}
    \begin{tabular}{cccccc}
        \hline
        Threshold & Gray value & Total events ($\geq$5 px) & Single-capillary events & Efficiency (\%) \\
        \hline
        $3\sigma$ & 0.0618 & 52,308 & 28,305 & 10.39 \\
        $4\sigma$ & 0.0783 & 51,323 & 27,803 & 10.20 \\
        $5\sigma$ & 0.0947 & 50,663 & 27,476 & 10.07 \\
        $6\sigma$ & 0.1112 & 50,325 & 27,359 & 10.00 \\
        \hline
    \end{tabular}
\end{table*}
To study the influence of the threshold on the detection efficiency, the analysis was repeated with $3\sigma$, $4\sigma$, $5\sigma$, and $6\sigma$ thresholds. Table~\ref{tab:threshold_scan} summarizes the results.

It can be seen that the detection efficiency decreases only slightly from 10.39\% at $3\sigma$ to 10.00\% at $6\sigma$, a decrease of about 0.4\% in absolute terms. This weak dependence on threshold indicates that the neutron-induced signal is well separated from the background noise in the capillary array detector, so that the efficiency is not sensitive to the exact choice of the threshold. The $5\sigma$ threshold is adopted in this work as a compromise between a low false-positive rate ($5.7\times10^{-7}$) and a good event retention.
The detection efficiency of 10.07\% is reasonable for the present capillary array detector. The main limiting factors include: (1) light loss during transmission in the capillaries, (2) the $5\sigma$ threshold which rejects a fraction of low-energy deposition events, and (3) the effective field of view which excludes edge regions with non-uniform background.  

It must be mentioned that the AmBe source emits gamma rays in addition to neutrons. The dominant high-energy gamma line is 4.44~MeV from the de-excitation of $^{12}\mathrm{C}^*$. Monte Carlo simulations using Geant4 show that 5~MeV neutrons and 4.44~MeV gamma rays behave very differently in a 50-$\mu$m capillary\cite{Zhang2022,Cai2019,Reichhart2012}. The recoil proton produced by neutron scattering has a short range and deposits nearly all its energy within the capillary, while the Compton electron produced by gamma rays has a much longer range and mostly escapes from the capillary, depositing only a small fraction of its energy. As a result, the fraction of gamma events exceeding the $5\sigma$ threshold is much lower than that of neutron events. Therefore, the $5\sigma$ threshold effectively suppresses the gamma background, and the measured efficiency is dominated by neutron events. Although a residual gamma contribution cannot be completely excluded, the majority of the detected events are still neutrons.
It should be noted that the AmBe source used in this experiment has an average neutron energy of approximately 5~MeV, whereas the detector is designed for 14~MeV fusion neutrons. Since the light output of the liquid scintillator depends on the energy deposited by the recoil proton, the detection efficiency measured with AmBe neutrons may differ from that for 14~MeV neutrons. In general, 14~MeV neutrons produce higher-energy recoil protons, leading to a larger light yield per event, which could result in a higher detection efficiency. Therefore, The efficiency measured with the AmBe source provides an experimental benchmark for the current detector prototype. It should be further noted that the quartz glass walls of the capillaries do not produce scintillation light. Only the liquid scintillator filling the capillary bores contributes to the signal. The effective area used in Eq.~(6) is the geometric area of the detector, which includes both the liquid scintillator and the glass walls. The inner diameter of each capillary is 50~$\mu$m, and the wall thickness is approximately 4~$\mu$m. The outer dimension of each hexagonal capillary cell is therefore about 58~$\mu$m, giving a fill factor of
\begin{equation}
f = \left( \frac{50}{58} \right)^2 \approx 0.743.
\end{equation}
That is, only about 74.3\% of the geometric area is occupied by the liquid scintillator. If the efficiency is normalized to the active liquid scintillator area instead of the geometric area, the corrected intrinsic efficiency becomes
\begin{equation}
\varepsilon_{\mathrm{liquid}} = \frac{\varepsilon}{f} = \frac{10.07\%}{0.743} \approx 13.55\%.
\end{equation}
This value represents the intrinsic detection efficiency of the EJ-309 liquid scintillator itself, excluding the inactive glass walls. In this work, the geometric-area-normalized efficiency of 10.07\% is reported as the detector-level efficiency, while the liquid-area-normalized value of 13.55\% is provided for comparison with the intrinsic response of the scintillator material.

\subsection{Single-Capillary Pulse Height Spectrum and Linearity}
\label{subsec:spectrum}
The pulse height spectrum of single-capillary events was extracted to characterize the detector response to individual neutron interactions. For each event classified as a single capillary (5--6 pixels), the total charge was obtained by integrating the background-subtracted gray values within a $9\times9$ pixel window centered at the event centroid. This window size was chosen to fully collect the scintillation light, including the low-intensity tails that spread beyond the core pixels.

To examine the influence of the threshold on the charge spectrum, the analysis was repeated with $3\sigma$, $4\sigma$, $5\sigma$, and $6\sigma$ thresholds. The resulting charge spectra were fitted with a Landau distribution function, from which the most probable value (MPV) and the width parameter were extracted. Table~\ref{tab:mpv_threshold} summarizes the fit results.
\begin{table}[htbp]
    \centering
    \caption{Landau fit results of the single-capillary charge spectrum under different selection thresholds.}
    \label{tab:mpv_threshold}
    \begin{tabular}{cccccc}
        \hline
        Threshold & Events & MPV & Width\\
        \hline
        $3\sigma$ & 28,305 & $0.1343 \pm 0.001$ & $0.0482 \pm 0.0006$  \\
        $4\sigma$ & 27,803 & $0.1333 \pm 0.001$ & $0.0485 \pm 0.0006$  \\
        $5\sigma$ & 27,476 & $0.1330 \pm 0.001$ & $0.0484 \pm 0.0006$  \\
        $6\sigma$ & 27,359 & $0.1317 \pm 0.001$ & $0.0476 \pm 0.0006$  \\
        \hline
    \end{tabular}
\end{table}

It can be seen that the MPV remains nearly constant across all four thresholds, varying only from 0.1343 ($3\sigma$) to 0.1317 ($6\sigma$), a relative change of less than 2\%. The width parameter $\sigma$ also remains stable at around 0.048. This weak dependence on the threshold confirms that the neutron-induced signal is well separated from the background noise. Even at a relatively low threshold of $3\sigma$, the charge spectrum retains the same shape and peak position, indicating that the choice of $5\sigma$ does not bias the charge measurement.
\begin{figure}[htbp]
    \centering
    \includegraphics[width=0.98\linewidth]{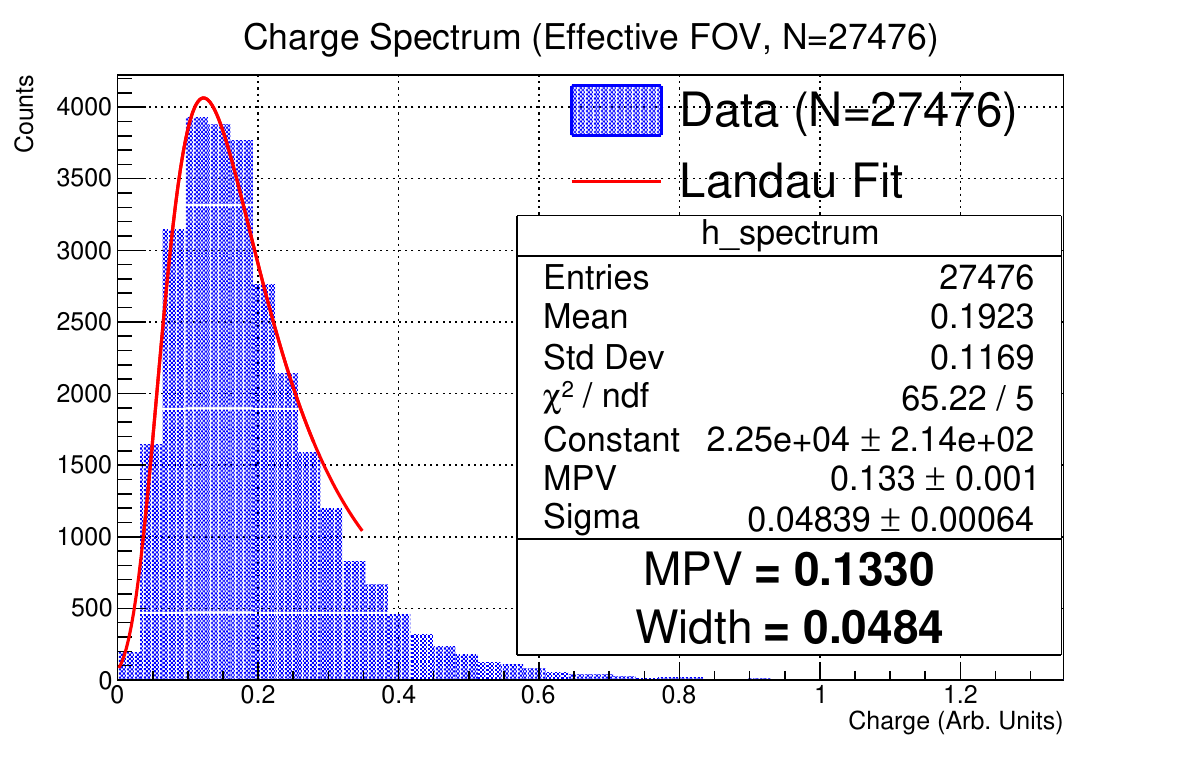}
    \caption{Charge spectrum of single-capillary events (5--6 pixels) in the effective field of view ($Y=1000$--$9000~\mu$m), obtained with the $5\sigma$ threshold. The red curve is the Landau fit with $\mathrm{MPV}=0.133\pm0.001$.}
    \label{fig:charge_single}
\end{figure}
Figure~\ref{fig:charge_single} shows the charge spectrum obtained with the $5\sigma$ threshold, together with the Landau fit. The fit yields $\mathrm{MPV} = 0.133 \pm 0.001$ and $\sigma = 0.0484 \pm 0.0006$, with a $\chi^2/\mathrm{ndf}$ of $65.22/5$. A total of 27,476 single-capillary events were used in the fit.
The Landau-like shape of the single-capillary charge spectrum is consistent with the energy deposition of recoil protons, which are produced by neutron elastic scattering on hydrogen and deposit nearly all their energy within one capillary. Gamma-induced events, in contrast, do not follow this distribution because the Compton electrons escape from the capillary and deposit only a small fraction of their energy.
The MPV of the single-capillary charge spectrum serves as a reference for the detector's energy response. The relatively large $\chi^2/\mathrm{ndf}$ may be attributed to the limited number of bins and the intrinsic asymmetry of the Landau distribution, but the MPV is robustly determined, as confirmed by the threshold scan.
\begin{figure*}[htbp]
    \centering
    \includegraphics[width=0.98\linewidth]{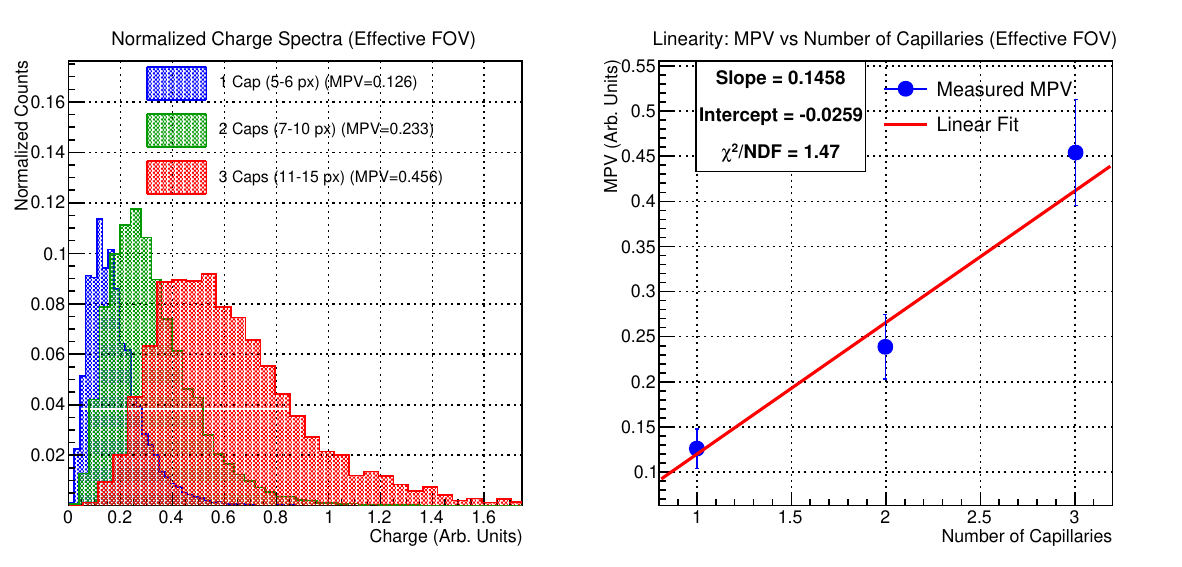}
    \caption{(Left) Normalized charge spectra of single-capillary (blue), two-capillary (green), and three-capillary (red) events, obtained with the $5\sigma$ threshold. (Right) MPV as a function of the number of capillaries, together with a linear fit.}
    \label{fig:mpv_linearity}
\end{figure*}

To study the linearity of the detector response, the charge spectra of single-capillary, two-capillary, and three-capillary events were extracted and fitted with a Landau distribution function. Figure~\ref{fig:mpv_linearity} shows the normalized charge spectra obtained with the $5\sigma$ threshold (left) and the MPV as a function of the number of capillaries (right).
The MPV values extracted from the fits are $0.126 \pm 0.001$ for single-capillary events, $0.233 \pm 0.002$ for two-capillary events, and $0.456 \pm 0.004$ for three-capillary events. The ratios between them are
\begin{equation}
\frac{\mathrm{MPV}(2\,\mathrm{cap})}{\mathrm{MPV}(1\,\mathrm{cap})} = 1.85, \qquad
\frac{\mathrm{MPV}(3\,\mathrm{cap})}{\mathrm{MPV}(1\,\mathrm{cap})} = 3.62.
\end{equation}
The ratio for two capillaries is close to the expected value of 2.0, with a deviation of about 7.5\%, indicating good linearity for the two-capillary case. The ratio for three capillaries, however, deviates from the expected value of 3.0 by about 20.7\%, revealing a clear nonlinearity for large events.

A linear fit was performed on the MPV values as a function of the number of capillaries,
\begin{equation}
\mathrm{MPV}(n) = k \cdot n + b,
\end{equation}
where $n$ is the number of capillaries. The slope $k$ represents the average charge contributed by each capillary, and the intercept $b$ reflects the baseline offset. The fit gives $k = 0.146$ (arb. units) and $b = -0.026$ (arb. units). The slope is consistent with the single-capillary MPV value (0.126) within about 12\%, and the small intercept indicates that the baseline is close to zero. These results suggest that the detector response is approximately linear for small events, but deviates from linearity for events involving three or more capillaries.

The deviation from linearity for three-capillary events is mainly attributed to the light spread effect. When three adjacent capillaries emit light simultaneously, the integration region defined by the connected pixels is larger than that for a single-capillary event. As a result, a larger fraction of the low-intensity light spread from neighboring capillaries is captured, which artificially increases the measured charge. This interpretation is consistent with the light spread analysis presented in Section~\ref{sec:laser}, where the light spot FWHM was found to be 90~$\mu$m even with the metal light absorber.

To verify that the observed nonlinearity originates from light spread rather than from an intrinsic nonlinear response of the detector, the same analysis was repeated with $3\sigma$, $4\sigma$, and $6\sigma$ thresholds. The MPV ratios vary only slightly across different thresholds, with $\mathrm{MPV}(2\,\mathrm{cap})/\mathrm{MPV}(1\,\mathrm{cap})$ ranging from 1.84 to 1.89 and $\mathrm{MPV}(3\,\mathrm{cap})/\mathrm{MPV}(1\,\mathrm{cap})$ ranging from 3.57 to 3.61. This weak dependence on threshold indicates that the nonlinearity is a robust feature of the detector response and is not an artifact of the threshold selection.

\subsection{Positioning Performance}
\label{subsec:positioning}
The positioning performance of the detector was characterized by the point spread function (PSF), which describes the spatial distribution of a point-like light source after passing through an imaging system~\cite{Goodman2005,Barrett1996,He2022}. For a single neutron event occurring in one capillary, the scintillation light forms a point-like source at the capillary end. After transmission through the capillary array and the optical system, the light is broadened into a finite spot on the camera sensor. The shape and width of this spot define the PSF of the detector~\cite{Disdier2004,Winick1986}.

The light spread in the capillary array was first studied in the laser experiment (Section~\ref{sec:laser}), where the laser beam was focused onto the capillary array and the resulting light spot was recorded in a single exposure. The FWHM of the full light spot was found to decrease from 260~$\mu$m (without absorber) to 90~$\mu$m (with absorber). This value characterizes the overall spatial extent of the light distribution on the array, including the low-intensity halo around the central peak.

\begin{figure*}[htbp]
    \centering
    \includegraphics[width=0.98\linewidth]{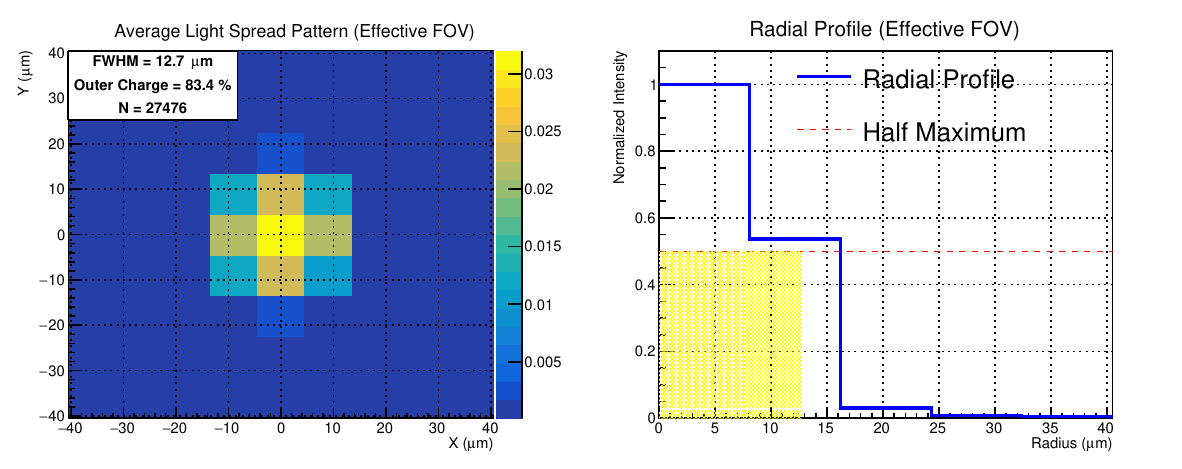}
    \caption{Average light spread pattern (left) and its radial profile (right) for single-capillary events. The radial profile is obtained by azimuthally averaging the 2D pattern around the centroid. The red dashed line indicates the half-maximum level, from which the FWHM of $12.70~\mu$m is extracted.}
    \label{fig:light_spread}
\end{figure*}

For the neutron experiment, the PSF core was extracted by aligning and superposing tens of thousands of single-capillary events within the effective field of view. By centering each event at its intensity-weighted centroid and averaging over all events, the low-intensity halo is smoothed out, and the sharp core of the PSF is revealed. Figure~\ref{fig:light_spread} shows the averaged light spread pattern and the corresponding radial profile.

The FWHM of the PSF core is measured to be
\begin{equation}
\mathrm{FWHM}_{\mathrm{core}} = 12.70~\mu\mathrm{m} \;(\approx 1.4\ \mathrm{pixels}),
\end{equation}
This value reflects the sharpness of the PSF core. The Outer Charge Fraction (OCF), defined as the fraction of charge deposited outside the central pixel within the $9\times9$ window, is approximately 83.4\%, indicating that most of the scintillation light from a single event is spread beyond the central pixel, forming a low-intensity halo around the central peak.

The centroid of each event was determined by the intensity-weighted centroid method,
\begin{equation}
\bar{x} = \frac{\sum_{i,j} g_{i,j} \cdot x_{i,j}}{\sum_{i,j} g_{i,j}}, \qquad
\bar{y} = \frac{\sum_{i,j} g_{i,j} \cdot y_{i,j}}{\sum_{i,j} g_{i,j}},
\end{equation}
where $g_{i,j}$ is the background-subtracted gray value at pixel $(i,j)$, and $(x_{i,j}, y_{i,j})$ is its physical coordinate. The statistical precision of the centroid, estimated from the PSF core width and the signal-to-noise ratio, is approximately $5.5~\mu$m ($1\sigma$). This precision reflects how accurately the center of an individual light spot can be located, but it does not represent the ability to distinguish two separate neutron events.

It is important to distinguish between the centroid positioning precision and the intrinsic position resolution of the detector. The centroid positioning precision of $5.5~\mu$m reflects the statistical accuracy with which the center of an individual light spot can be located. In contrast, the intrinsic position resolution (the minimum distance at which two neutron events can be distinguished) is limited by the capillary pitch of 54~$\mu$m. To verify this, the nearest-neighbor distance between independent events within the same frame was analyzed. A total of 30,118 events have a nearest neighbor within 500~$\mu$m. The nearest-neighbor distance distribution exhibits a prominent peak near 54~$\mu$m, which is consistent with the capillary pitch. Given the pixel size of 9~$\mu$m, we attribute this peak to the intrinsic position resolution limit of the detector, determined by the capillary pitch of 54~$\mu$m.

To verify that the measured PSF core width is an intrinsic property of the detector, the analysis was repeated with $3\sigma$, $4\sigma$, $5\sigma$, and $6\sigma$ thresholds. The FWHM remains stable at $12.76 \pm 0.05~\mu$m (variation less than 0.5\%) across all thresholds, confirming that the core width is determined by the optical system and is independent of the threshold selection. It should also be emphasized that the outer charge originates from the spatial spread of light from the same neutron event, not from crosstalk between adjacent capillaries. Since the light spot is spatially continuous and centered on a single event, all the charge within the $9\times9$ window belongs to that event. The measured OCF therefore reflects the intrinsic optical spread of the detector, rather than any inter-channel interference.

\section{Conclusion}
\label{sec:summary}
In this work, a neutron detector based on a hexagonal capillary array filled with EJ-309 liquid scintillator was developed and characterized. The detector uses capillaries with an inner diameter of 50~$\mu$m, and the scintillation light is recorded by a camera with 9~$\mu$m pixels. The main results are summarized as follows.

The light spread mechanism was investigated through laser experiments. With a 532~nm green laser, the FWHM of the full light spot was measured to be 260~$\mu$m without a light absorber and 90~$\mu$m with a metal light absorber, demonstrating that the absorber effectively suppresses lateral light spread. For neutron irradiation experiments with an AmBe source, an effective field of view and a $5\sigma$ threshold were established from 11 background frames. Based on the geometric area, the intrinsic detection efficiency was determined to be $10.07\% \pm 1.26\%~(\mathrm{RMS}) \pm 1.43\%~(\mathrm{syst.})$. After correcting for the fill factor of the liquid scintillator ($f \approx 0.743$), the intrinsic efficiency of the EJ-309 liquid scintillator itself was estimated to be about 13.55\%. The efficiency varied by only 0.4\% when the threshold was changed from $3\sigma$ to $6\sigma$, indicating that the signal is well separated from the background noise.

The pulse height spectrum of single-capillary events follows a Landau distribution with a MPV of $0.133 \pm 0.001$ when using a $9\times9$ pixel integration window. The central pixel accounts for only 16.6\% of the total charge, while the OCF is about 83.4\%, confirming that the scintillation light spreads significantly beyond the central pixel. Linearity analysis shows that the MPV ratio for two capillaries is 1.85, close to the expected value of 2.0, while the ratio for three capillaries is 3.62, deviating from 3.0 due to additional capture of spread light by the larger integration window. The positioning performance was characterized by the PSF. The core FWHM is 12.70~$\mu$m, and the centroid positioning precision is approximately 5.5~$\mu$m. The intrinsic position resolution of the detector is limited by the capillary pitch to about 54~$\mu$m. The PSF core width remains stable within 0.5\% across thresholds from $3\sigma$ to $6\sigma$.

The significance of this work lies in several aspects. First, the lateral light spread in a capillary liquid scintillator detector was experimentally quantified, showing that it is a continuous, low-intensity distribution rather than discrete crosstalk. Second, a metal light absorber coated on the capillary outer walls effectively suppresses this spread, providing a practical solution that can be integrated into the existing fabrication process and improves multi-capillary linearity. Third, we clarified three distinct spatial performance metrics: centroid positioning precision, PSF core width, and intrinsic position resolution. This framework provides concrete engineering guidance: for flux mapping, the signal-to-noise ratio should be enhanced and the centroid algorithm optimized; for high-resolution imaging, the capillary pitch should be reduced (e.g., by decreasing the inner diameter or wall thickness), though this must be balanced against detection efficiency; for multi-capillary events, reflective or absorbing coatings between capillaries are essential to suppress light spread and maintain linearity; and for further improving the centroid precision, the optical system can be optimized to reduce the PSF core width. Future detector designs should optimize these measures according to the specific application requirements.

\section{Acknowledgements}
\label{sec:ACK}
This work is supported by the Basic Research Program of Shenzhen (JCYJ20190807155418935), Natural Science Foundation of Top Talent of SZTU (grant no. 20200206), National Natural Science Foundation of China (12405247), and Guangdong Provincial Department of Young Innovative Talents Project (2024KQNCX038).


\begin{thebibliography}{99}
\bibitem{Frenje2020}
J. A. Frenje, Nuclear diagnostics for inertial confinement fusion (ICF) plasmas, Plasma Phys. Control. Fusion 62 (2020) 044pp.
\href{http://dx.doi.org/10.1088/1361-6587/ab5137}{http://dx.doi.org/10.1088/1361-6587/ab5137}

\bibitem{Caillaud2012a}
T. Caillaud, et al., Development of the large neutron imaging system for inertial confinement fusion experiments, Rev. Sci. Instrum. 83 (2012) 033502.
\href{http://dx.doi.org/10.1063/1.3689768}{http://dx.doi.org/10.1063/1.3689768}

\bibitem{Caillaud2012b}
T. Caillaud, et al., A new compact, high sensitivity neutron imaging system, Rev. Sci. Instrum. 83 (2012) 10E131.
\href{http://dx.doi.org/10.1063/1.4739314}{http://dx.doi.org/10.1063/1.4739314}

\bibitem{Miyanaga1990}
N. Miyanaga, et al., Neutron penumbral imaging of laser fusion targets, Rev. Sci. Instrum. 61 (1990) 3230.
\href{http://dx.doi.org/10.1063/1.1141645}{http://dx.doi.org/10.1063/1.1141645}

\bibitem{He2022}
Z. He, N. Huang, P. Wang, et al., Spatial resolution and image processing for pinhole camera-based X-ray fluorescence imaging: a simulation study, Nucl. Sci. Tech. 33 (2022) 64.
\href{https://doi.org/10.1007/s41365-022-01036-8}{https://doi.org/10.1007/s41365-022-01036-8}

\bibitem{Guler2012}
N. Guler, et al., Simultaneous usage of pinhole and penumbral apertures for imaging small scale neutron sources from inertial confinement fusion experiments, Rev. Sci. Instrum. 83 (2012) 10D316.
\href{http://dx.doi.org/10.1063/1.4746745}{http://dx.doi.org/10.1063/1.4746745}

\bibitem{Isabelle2010}
T. Isabelle, et al., Alignment effects on a neutron imaging system using coded apertures, Rev. Sci. Instrum. 81 (2010) 033304.
\href{http://dx.doi.org/10.1063/1.3331494}{http://dx.doi.org/10.1063/1.3331494}

\bibitem{Wilke2008}
M. D. Wilke, et al., The National Ignition Facility neutron imaging system, Rev. Sci. Instrum. 79 (2008) 10E529.
\href{http://dx.doi.org/10.1063/1.2987984}{http://dx.doi.org/10.1063/1.2987984}

\bibitem{Ma2025}
Y.H. Ma, B. Tang, W. Yin, et al., Single-neutron super-resolution imaging based on neutron capture event detection and reconstruction, Nucl. Sci. Tech. 36 (2025) 114.
\href{https://doi.org/10.1007/s41365-025-01668-6}{https://doi.org/10.1007/s41365-025-01668-6}

\bibitem{Lerche2003}
R. A. Lerche, et al., Neutron images recorded with high-resolution bubble detectors, Rev. Sci. Instrum. 74 (2003) 1709--1712.
\href{http://dx.doi.org/10.1063/1.1534931}{http://dx.doi.org/10.1063/1.1534931}

\bibitem{Mor2012}
I. Mor, et al., Fast-neutron imaging spectrometer based on liquid scintillator loaded capillaries, J. Instrum. 7 (2012) C04021.
\href{http://dx.doi.org/10.1088/1748-0221/7/04/C04021}{http://dx.doi.org/10.1088/1748-0221/7/04/C04021}

\bibitem{Li2022}
Y.T. Li, W.P. Lin, B.S. Gao, et al., Development of a low-background neutron detector array, Nucl. Sci. Tech. 33 (2022) 41.
\href{https://doi.org/10.1007/s41365-022-01030-0}{https://doi.org/10.1007/s41365-022-01030-0}

\bibitem{Chen2022}
W.K. Chen, L.Q. Hu, G.Q. Zhong, et al., Optimization study and design of scintillating fiber detector for D-T neutron measurements on EAST with Geant4, Nucl. Sci. Tech. 33 (2022) 139.
\href{https://doi.org/10.1007/s41365-022-01123-w}{https://doi.org/10.1007/s41365-022-01123-w}

\bibitem{Zhang2023}
Y.Q. Zhang, L.Q. Hu, G.Q. Zhong, et al., Development of a high-speed digital pulse signal acquisition and processing system based on MTCA for liquid scintillator neutron detector on EAST, Nucl. Sci. Tech. 34 (2023) 150.
\href{https://doi.org/10.1007/s41365-023-01318-9}{https://doi.org/10.1007/s41365-023-01318-9}

\bibitem{Disdier2004}
L. Disdier, et al., Capillary detector with deuterated scintillator for inertial confinement fusion neutron images, Rev. Sci. Instrum. 75 (2004) 2134--2139.
\href{http://dx.doi.org/10.1063/1.1755443}{http://dx.doi.org/10.1063/1.1755443}

\bibitem{Disdier2006}
L. Disdier, et al., Inertial confinement fusion neutron images, Phys. Plasmas 13 (2006) 056317.
\href{http://dx.doi.org/10.1063/1.2174828}{http://dx.doi.org/10.1063/1.2174828}

\bibitem{Zhao2023}
Q. Zhao, Y.B. Nie, Y.Y. Ding, et al., Measurement and simulation of the leakage neutron spectra from Fe spheres bombarded with 14 MeV neutrons, Nucl. Sci. Tech. 34 (2023) 182.
\href{https://doi.org/10.1007/s41365-023-01329-6}{https://doi.org/10.1007/s41365-023-01329-6}

\bibitem{Ding2024}
Y.Y. Ding, Y.B. Nie, Y. Zhang, et al., Benchmark experiment on slab $^{238}$U with D-T neutrons for validation of evaluated nuclear data, Nucl. Sci. Tech. 35 (2024) 29.
\href{https://doi.org/10.1007/s41365-024-01386-5}{https://doi.org/10.1007/s41365-024-01386-5}

\bibitem{Chen2019}
Z. Chen, et al., Design of neutron imaging aperture for inertial confinement fusion in laser fusion research center, J. Instrum. 14 (2019) C11007.
\href{http://dx.doi.org/10.1088/1748-0221/14/11/C11007}{http://dx.doi.org/10.1088/1748-0221/14/11/C11007}

\bibitem{Vogel2014}
P. L. Volegov, et al., Self characterization of a coded aperture array for neutron source imaging, Rev. Sci. Instrum. 85 (2014) 123506.
\href{http://dx.doi.org/10.1063/1.4902978}{http://dx.doi.org/10.1063/1.4902978}

\bibitem{Zhang2022}
C. Zhang, et al.,
Simulation of a micron resolution capillary liquid scintillation detector for 14 MeV fusion neutrons,
Applied Radiation and Isotopes 189 (2022) 110424.
\href{https://doi.org/10.1016/j.apradiso.2022.110424}{https://doi.org/10.1016/j.apradiso.2022.110424}

\bibitem{Song2018}
H.H. Song, Y.G. Yuan, T.P. Peng, et al., Optimization study on neutron spectrum unfolding based on the least-squares method, Nucl. Sci. Tech. 29 (2018) 118.
\href{https://doi.org/10.1007/s41365-018-0454-5}{https://doi.org/10.1007/s41365-018-0454-5}

\bibitem{Li1989a}
H. Li, P. Dao, R. Jayakumar, Improvements and systolic implementation of the Hough transformation for straight line detection, Pattern Recognit. 22 (1989) 697--706.
\href{http://dx.doi.org/10.1016/0031-3203(89)90006-X}{http://dx.doi.org/10.1016/0031-3203(89)90006-X}


\bibitem{Fernandes2008}
A. L. Fernandes, M. M. Oliveira, Real-time line detection through an improved Hough transform voting scheme, Pattern Recognit. 41 (2008) 299--314.
\href{http://dx.doi.org/10.1016/j.patcog.2007.04.003}{http://dx.doi.org/10.1016/j.patcog.2007.04.003}

\bibitem{He2019}
T. He, P. Zheng, J. Xiao, Measurement of the prompt neutron spectrum from thermal-neutron-induced fission in U-235 using the recoil proton method, Nucl. Sci. Tech. 30 (2019) 112.
\href{https://doi.org/10.1007/s41365-019-0633-z}{https://doi.org/10.1007/s41365-019-0633-z}

\bibitem{Song2016}
Y. Song, et al., Monte Carlo simulation of a very high resolution thermal neutron detector composed of glass scintillator microfibers, Appl. Radiat. Isot. 108 (2016) 100--107.
\href{http://dx.doi.org/10.1016/j.apradiso.2015.12.035}{http://dx.doi.org/10.1016/j.apradiso.2015.12.035}

\bibitem{Song2020}
Z. Song, et al., A simulation study of a high-resolution fast neutron imaging detector based on liquid scintillator loaded capillaries, Radiat. Detect. Technol. Methods 4 (2020) 152--160.
\href{http://dx.doi.org/10.1007/s41605-020-00164-2}{http://dx.doi.org/10.1007/s41605-020-00164-2}

\bibitem{Liu2023}
Y. Liu, T.F. Zhu, Z. Luo, et al., First-order primal-dual algorithm for sparse-view neutron computed tomography-based three-dimensional image reconstruction, Nucl. Sci. Tech. 34 (2023) 118.
\href{https://doi.org/10.1007/s41365-023-01258-4}{https://doi.org/10.1007/s41365-023-01258-4}

\bibitem{Kang2025}
M.X. Kang, J.Z. Zhang, H.Y. Wu, et al., Commissioning of the fast neutron detector array at China Institute of Atomic Energy, Nucl. Sci. Tech. 36 (2025) 86.
\href{https://doi.org/10.1007/s41365-025-01649-9}{https://doi.org/10.1007/s41365-025-01649-9}

\bibitem{Enqvist2013}
A. Enqvist, et al., Neutron light output response and resolution functions in EJ-309 liquid scintillation detectors, Nucl. Instrum. Methods Phys. Res. A 715 (2013) 79--86.
\href{http://dx.doi.org/10.1016/j.nima.2013.03.032}{http://dx.doi.org/10.1016/j.nima.2013.03.032}

\bibitem{Norsworthy2018}
M. A. Norsworthy, et al., Light output response of EJ-309 liquid organic scintillator to 2.86--3.95 MeV carbon recoil ions, Nucl. Instrum. Methods Phys. Res. A 884 (2018) 82--91.
\href{http://dx.doi.org/10.1016/j.nima.2017.11.084}{http://dx.doi.org/10.1016/j.nima.2017.11.084}

\bibitem{Tretyak2010}
V. I. Tretyak, Semi-empirical calculation of quenching factors for ions in scintillators, Astropart. Phys. 33 (2010) 40--53.
\href{http://dx.doi.org/10.1016/j.astropartphys.2009.11.002}{http://dx.doi.org/10.1016/j.astropartphys.2009.11.002}

\bibitem{Swiderski2012}
L. Swiderski, et al., Electron response of some low-Z scintillators in wide energy range, J. Instrum. 7 (2012) P06011.
\href{http://dx.doi.org/10.1088/1748-0221/7/06/P06011}{http://dx.doi.org/10.1088/1748-0221/7/06/P06011}

\bibitem{NorthNightVision}
North Night Vision Technology (Nanjing) Research Institute Co., Ltd., \url{http://bfys.norincogroup.com.cn/index.html}.

\bibitem{Samet1986}
H. Samet, An improved approach to connected component labeling of images, in: Proceedings of the IEEE Computer Society Conference on Computer Vision and Pattern Recognition (CVPR), 1986, pp. 312--318.
\href{https://www.cs.umd.edu/~hjs/pubs/SametCVPR86c.pdf}{https://www.cs.umd.edu/~hjs/pubs/SametCVPR86c.pdf}

\bibitem{Zhou2011}
L. Zhou, Y. Ye, L. Liu, et al., Parallel connected component detection algorithm for multi-core based on BFS, Opto-Electron. Eng. 38 (2011) 86.
\href{https://doi.org/10.3969/j.issn.1003-501X.2011.07.016}{https://doi.org/10.3969/j.issn.1003-501X.2011.07.016}

\bibitem{Cai2019}
J.L. Cai, D.W. Li, P.L. Wang, et al., Fast pulse sampling module for real-time neutron--gamma discrimination, Nucl. Sci. Tech. 30 (2019) 84.
\href{https://doi.org/10.1007/s41365-019-0595-1}{https://doi.org/10.1007/s41365-019-0595-1}

\bibitem{Reichhart2012}
L. Reichhart, D.Yu. Akimov, H.M. Ara\'ujo, et al., Quenching factor for low-energy nuclear recoils in a plastic scintillator, Phys. Rev. C 85 (2012) 065801.
\href{https://doi.org/10.1103/PhysRevC.85.065801}{https://doi.org/10.1103/PhysRevC.85.065801}

\bibitem{Goodman2005}
J. W. Goodman, Introduction to Fourier Optics, 3rd ed., Roberts and Company Publishers, 2005.

\bibitem{Barrett1996}
H. H. Barrett, W. Swindell, Radiological Imaging: The Theory of Image Formation, Detection, and Processing, Academic Press, 1981.

\bibitem{Winick1986}
K. A. Winick, Cram\'er-Rao lower bounds on the performance of charge-coupled-device optical position estimators, J. Opt. Soc. Am. A 3 (1986) 1809--1815.
\href{https://doi.org/10.1364/JOSAA.3.001809}{https://doi.org/10.1364/JOSAA.3.001809}


\end{thebibliography}
\end{document}